\documentclass{webofc}
\usepackage[varg]{txfonts}   
\begin{document}
\title{Resonances in a sudden chemical freeze-out model}
%
%

\author{\firstname{Viktor} \lastname{Begun}\inst{1}\fnsep\thanks{\email{viktor.begun@gmail.com}} 
}

\institute{Warsaw University of Technology, Faculty of Physics, Koszykowa 75, 00-662 Warsaw Poland
          }

\abstract{The prediction for $p_T$ spectra of various resonances produced in Pb+Pb collisions at 2.76 TeV at the LHC in equilibrium and non-equilibrium models is made. It includes the $\eta$, $\rho$(770), $\Sigma$(1385), $\Lambda$(1520), and $\Xi$(1530). The apparent differences may allow to distinguish between the models. }
\maketitle
%
A chemical non-equilibrium model~\cite{Petran:2013lja} with a single freeze-out~\cite{Chojnacki:2011hb} appeared to be rather successful in describing the LHC ALICE data at 2.76 TeV for various particles~\cite{Begun:2013nga,Begun:2014rsa}. In this model the mean {\it multiplicities} are described with the use of four thermodynamic parameters: temperature $T$, volume $V$, and two non-equilibrium parameters - $\gamma_s$ and $\gamma_q$.
It fixes the area under the curve for the $p_T$ spectra. 
The {\it form} of the spectra is best reproduced by the Hubble-like single freeze-out hyper-surface.
Then, the {\it slopes} of the spectra are described with only one extra parameter - the ratio of the freeze-out time $\tau_f$ to the freeze-out radius $r_f$, because their combination, the cylinder of volume $\pi*r^2*\tau$, is equal to the volume $V$, which is determined from multiplicities on the previous step.

It appears that the $p_T$ spectra of pions, kaons, protons, $K^*(892)^0$, and $\phi(1020)$ are described by the same parameters in the single freeze-out model~\cite{Begun:2013nga,Begun:2014rsa}. 
This is very surprising for the $K^*(892)^0$ and the $\phi(1020)$, because the first one is short living, while the second one is long living. The description of both of them may question the necessity of the long re-scattering phase, which is also successfully used to describe the ALICE data~\cite{Knospe:2015nva}. It may also indicate that the non-equilibrium, as implemented in~\cite{Begun:2013nga,Begun:2014rsa}, may effectively include the re-scattering, because $\gamma_q$ and $\gamma_s$ are equivalent to non-equilibrium chemical potentials for each particle, see~\cite{Begun:2013nga,Begun:2014rsa}. 
It is important to differentiate between the equilibrium with the re-scattering, and the single sudden freeze-out in the non-equilibrium, because the non-equilibrium also leads to pion condensation~\cite{Begun:2015ifa,Begun:2016cva}.

\begin{figure*}
\centering
 \includegraphics[width=0.47\textwidth,height=0.39\textwidth]{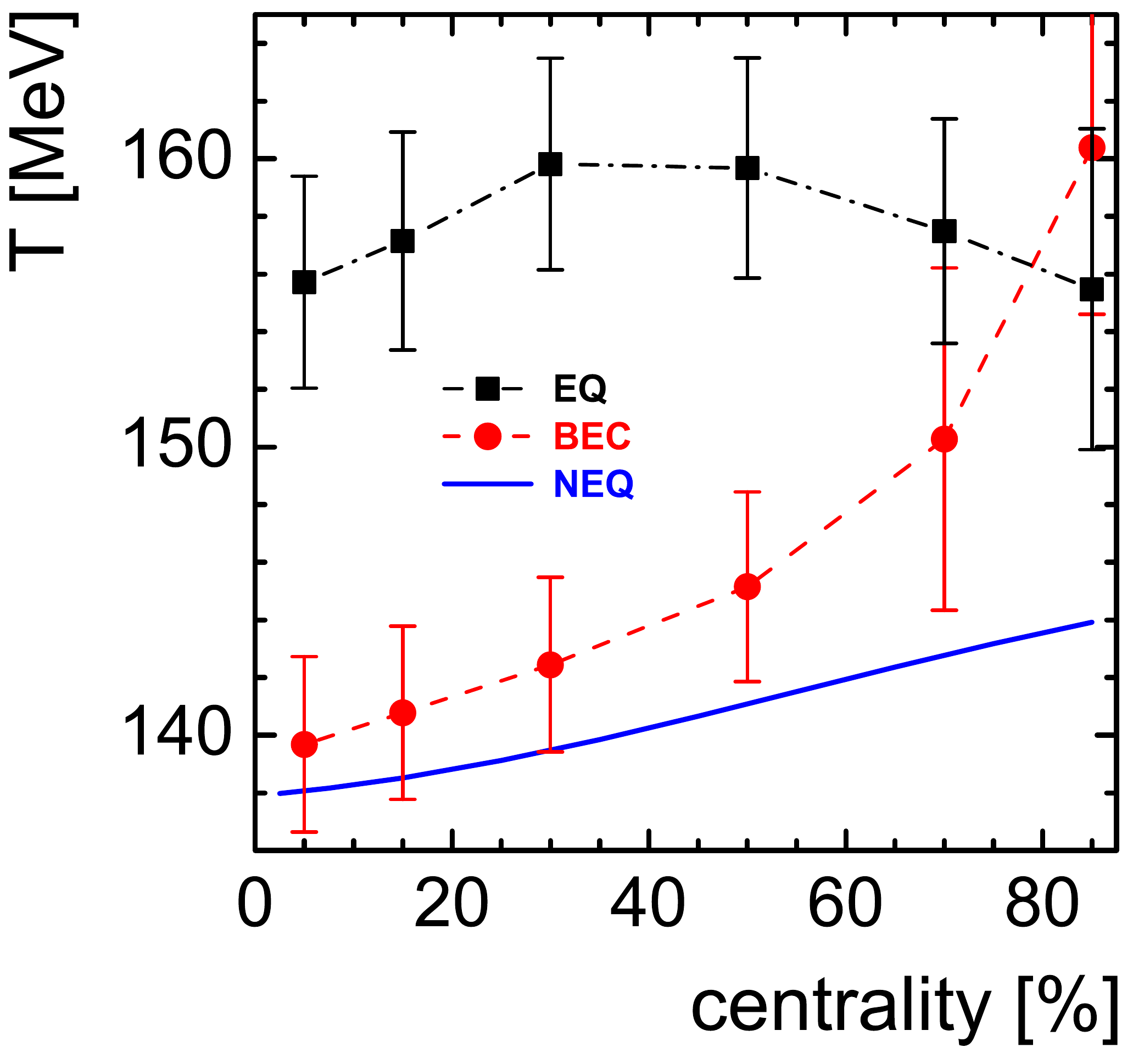}\hspace{0.05\textwidth}
 \includegraphics[width=0.47\textwidth,height=0.39\textwidth]{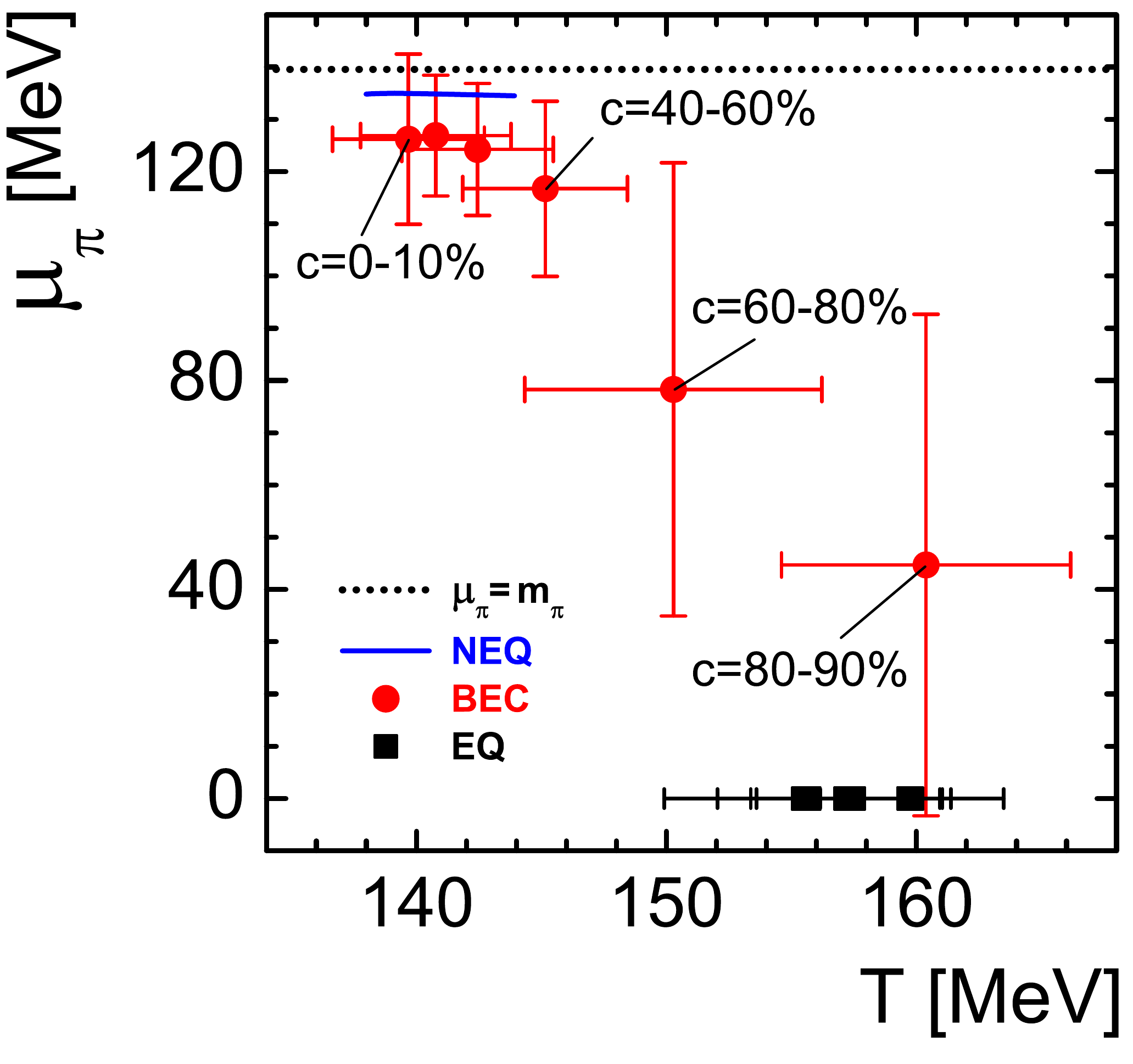}\vspace{-0.1cm}
\caption{
Temperature and non-equilibrium pion chemical potential, $\mu_{\pi}=2T\ln\gamma_q$, in Pb+Pb at 2.76 TeV~\cite{Begun:2015ifa,Begun:2016cva}.}
\vspace{-0.2cm}\label{fig-1}  
\end{figure*}

\begin{figure*}
\centering
 \includegraphics[width=0.47\textwidth,height=0.39\textwidth]{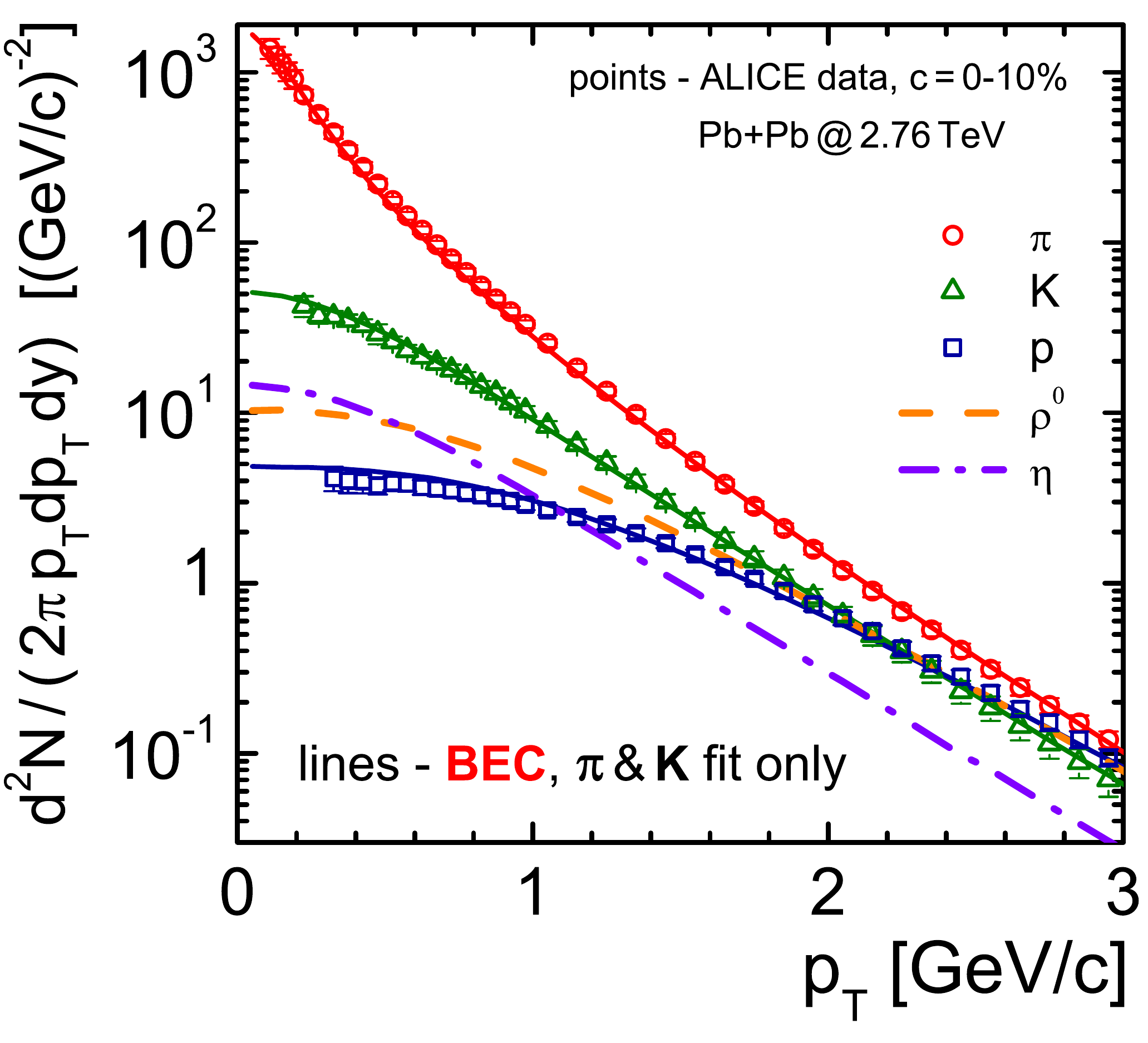}\hspace{0.05\textwidth}
 \includegraphics[width=0.47\textwidth,height=0.39\textwidth]{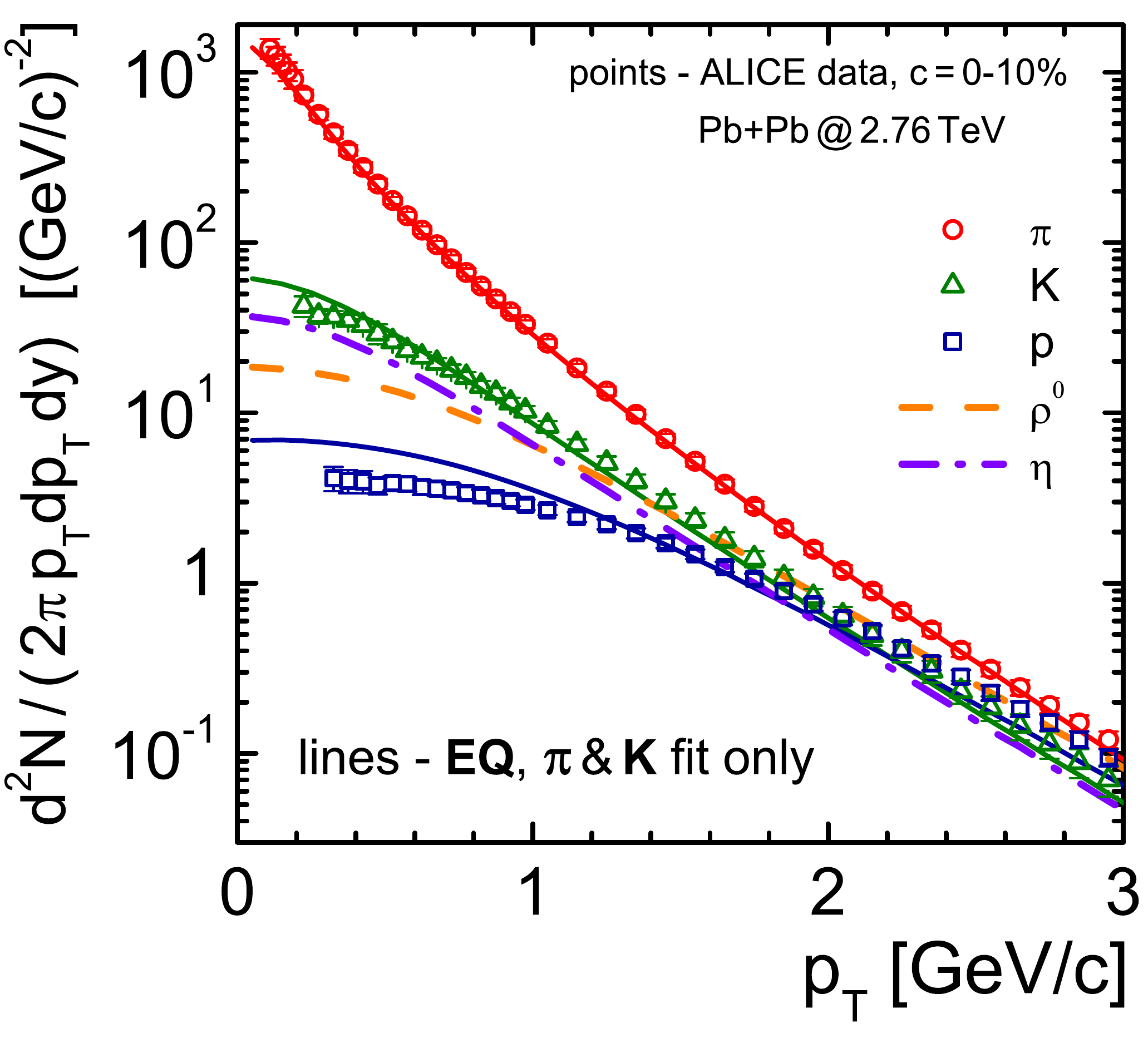}\vspace{-0.1cm}
\caption{The fit of pions and kaons, and the prediction for protons, $\rho$ and $\eta$ in 0-10\% centrality window.}
\vspace{-0.2cm}\label{fig-2}  
\end{figure*}

\begin{figure*}
\centering
 \includegraphics[width=0.47\textwidth,height=0.39\textwidth]{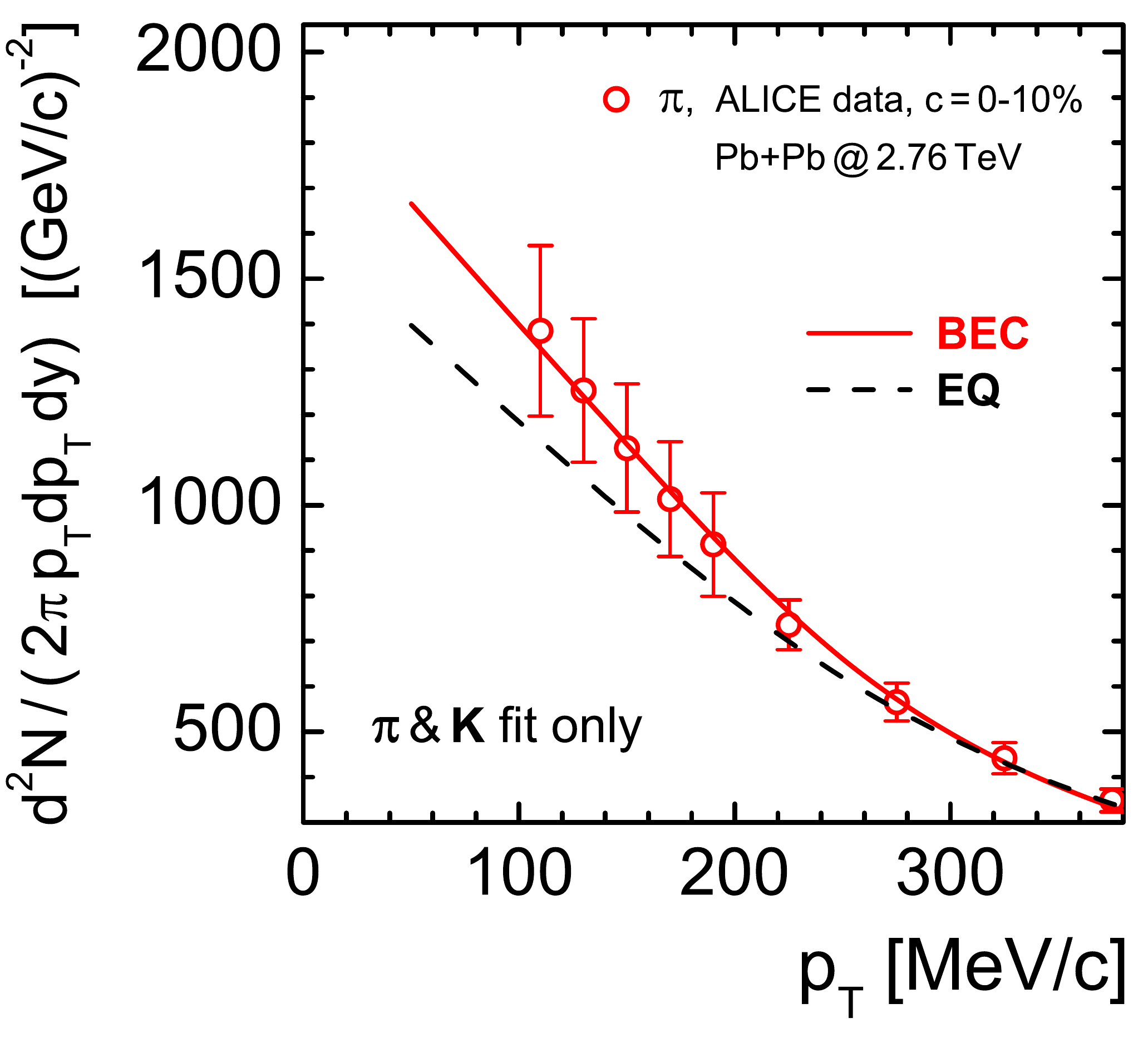}\hspace{0.05\textwidth}
 \includegraphics[width=0.47\textwidth,height=0.39\textwidth]{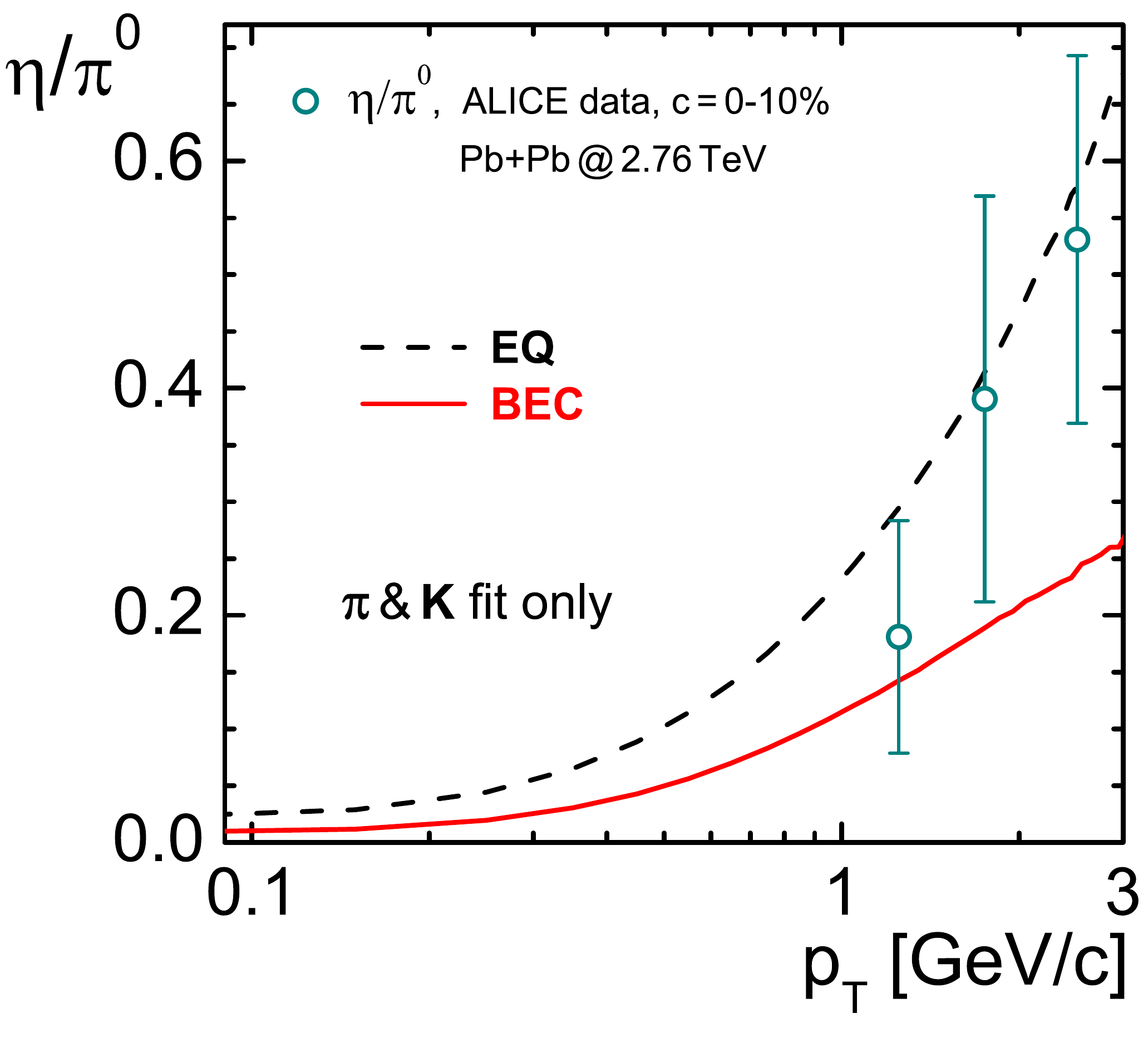}\vspace{-0.1cm}
\caption{Low $p_T$ charged pions and $\eta/\pi^0$ in EQ and BEC.}
\vspace{-0.2cm}\label{fig-3}  
\end{figure*}

\begin{figure*}
\centering
 \includegraphics[width=0.47\textwidth,height=0.39\textwidth]{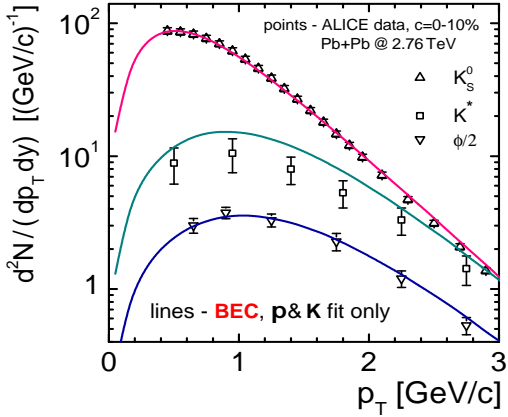}\hspace{0.05\textwidth}
 \includegraphics[width=0.47\textwidth,height=0.39\textwidth]{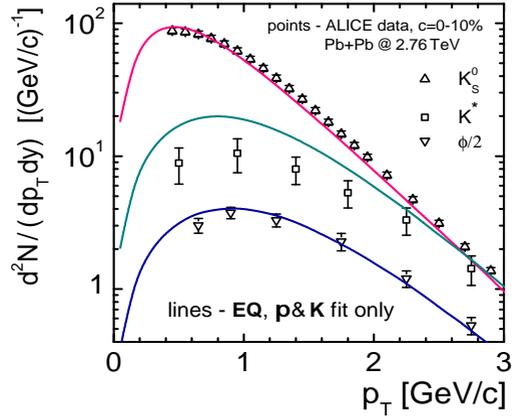}\vspace{-0.1cm}
\caption{Strange mesons obtained for the hypersurface, which was fitted to pions and kaons only.}
\vspace{-0.2cm}\label{fig-4}  
\end{figure*}

\begin{figure*}
\centering
 \includegraphics[width=0.47\textwidth,height=0.39\textwidth]{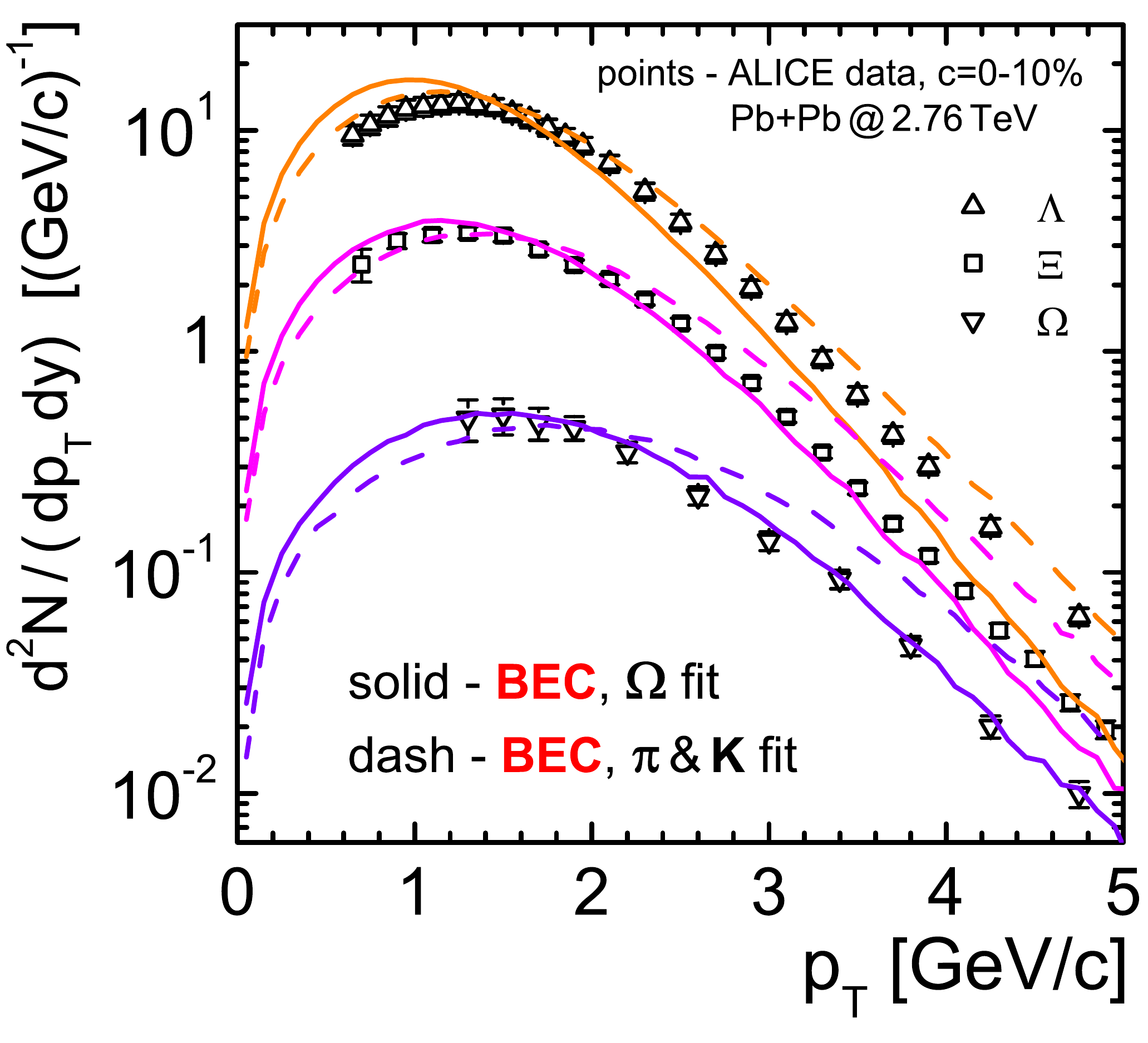}\hspace{0.05\textwidth}
 \includegraphics[width=0.47\textwidth,height=0.39\textwidth]{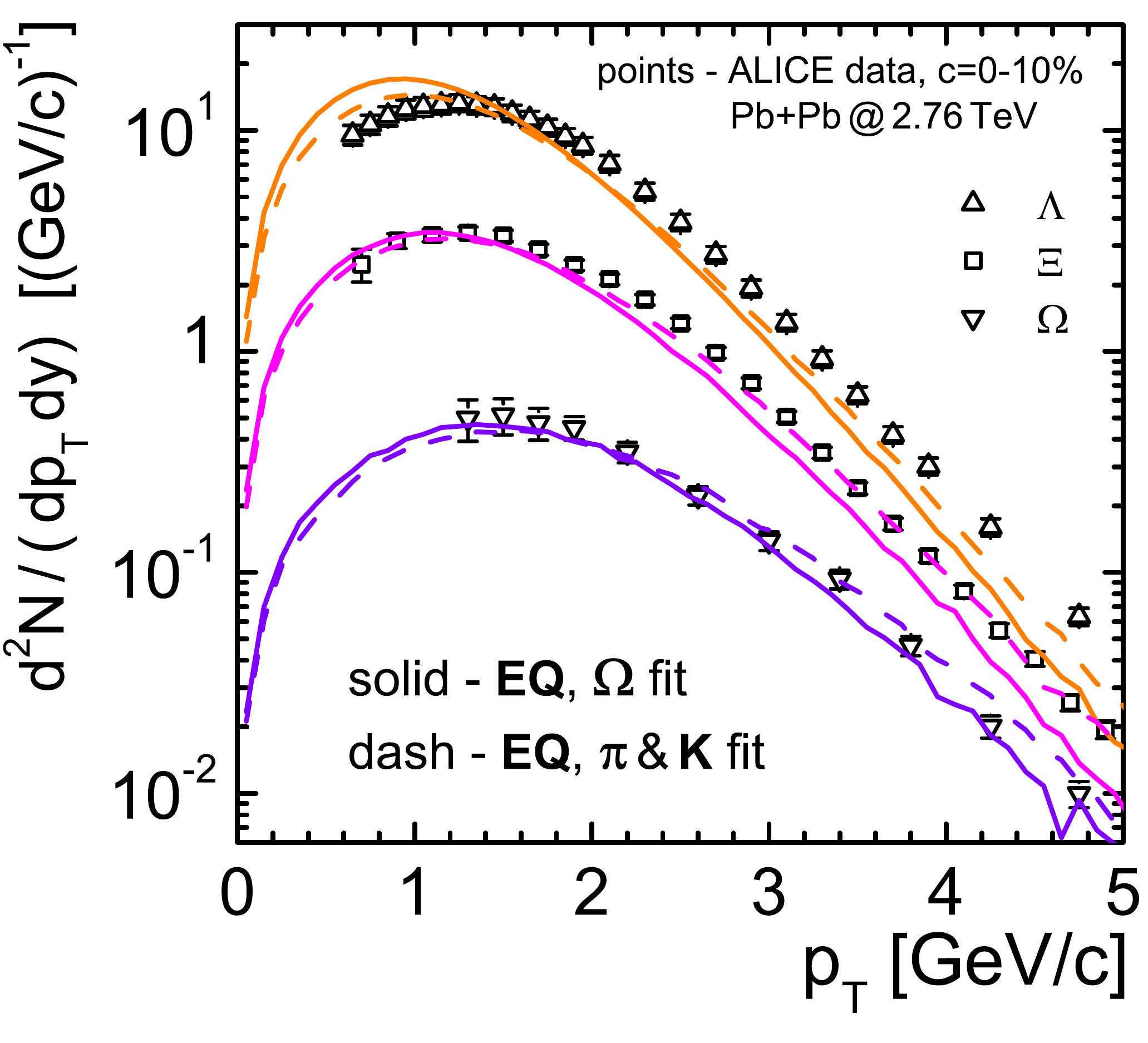}\vspace{-0.1cm}
\caption{Strange baryons obtained for the hypersurface that was fitted either to pions and kaons, or to $\Omega$ baryons.}
\vspace{-0.2cm}\label{fig-5}  
\end{figure*}

\begin{figure*}
\centering
 \includegraphics[width=0.47\textwidth,height=0.39\textwidth]{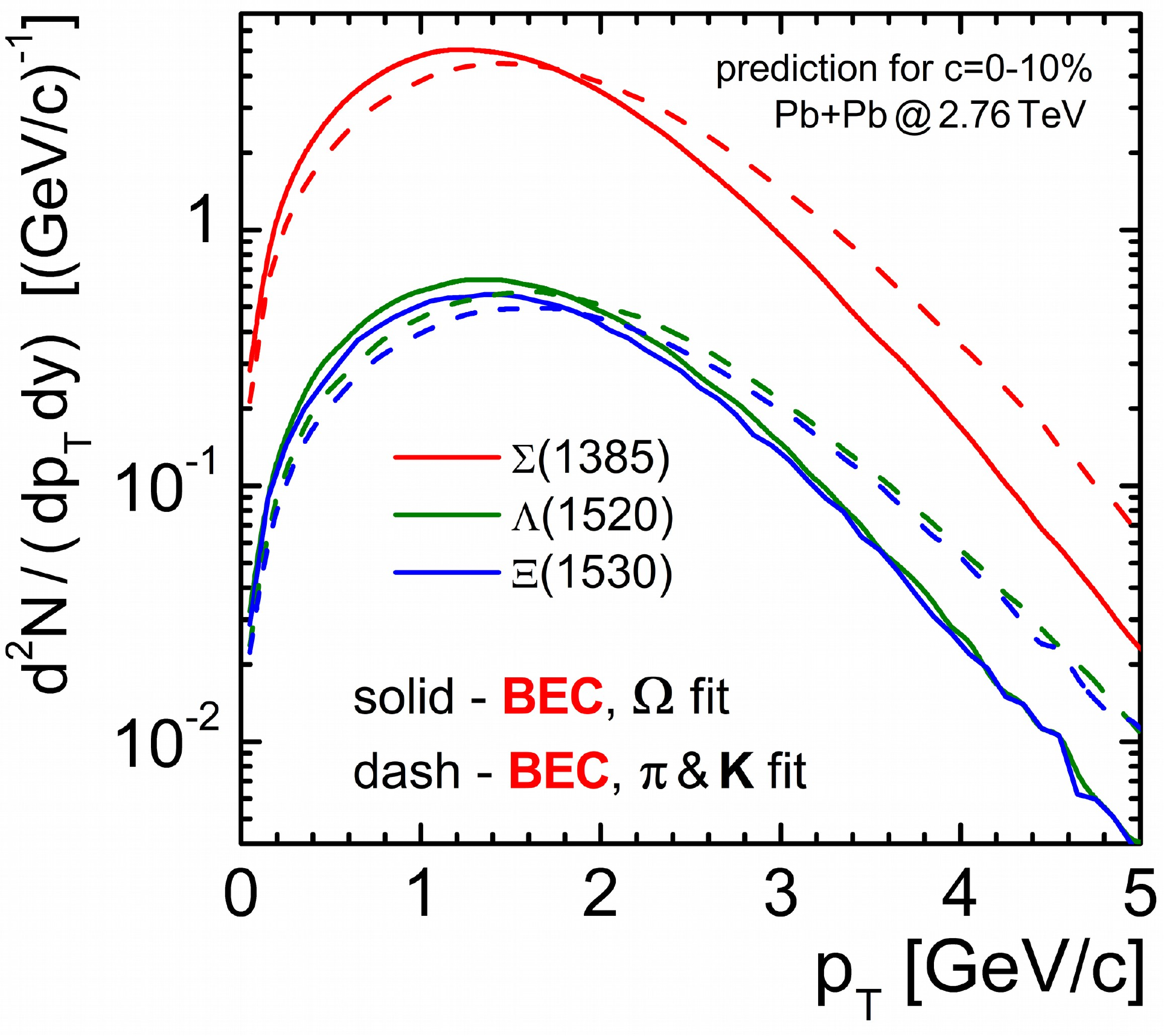}\hspace{0.05\textwidth}
 \includegraphics[width=0.47\textwidth,height=0.39\textwidth]{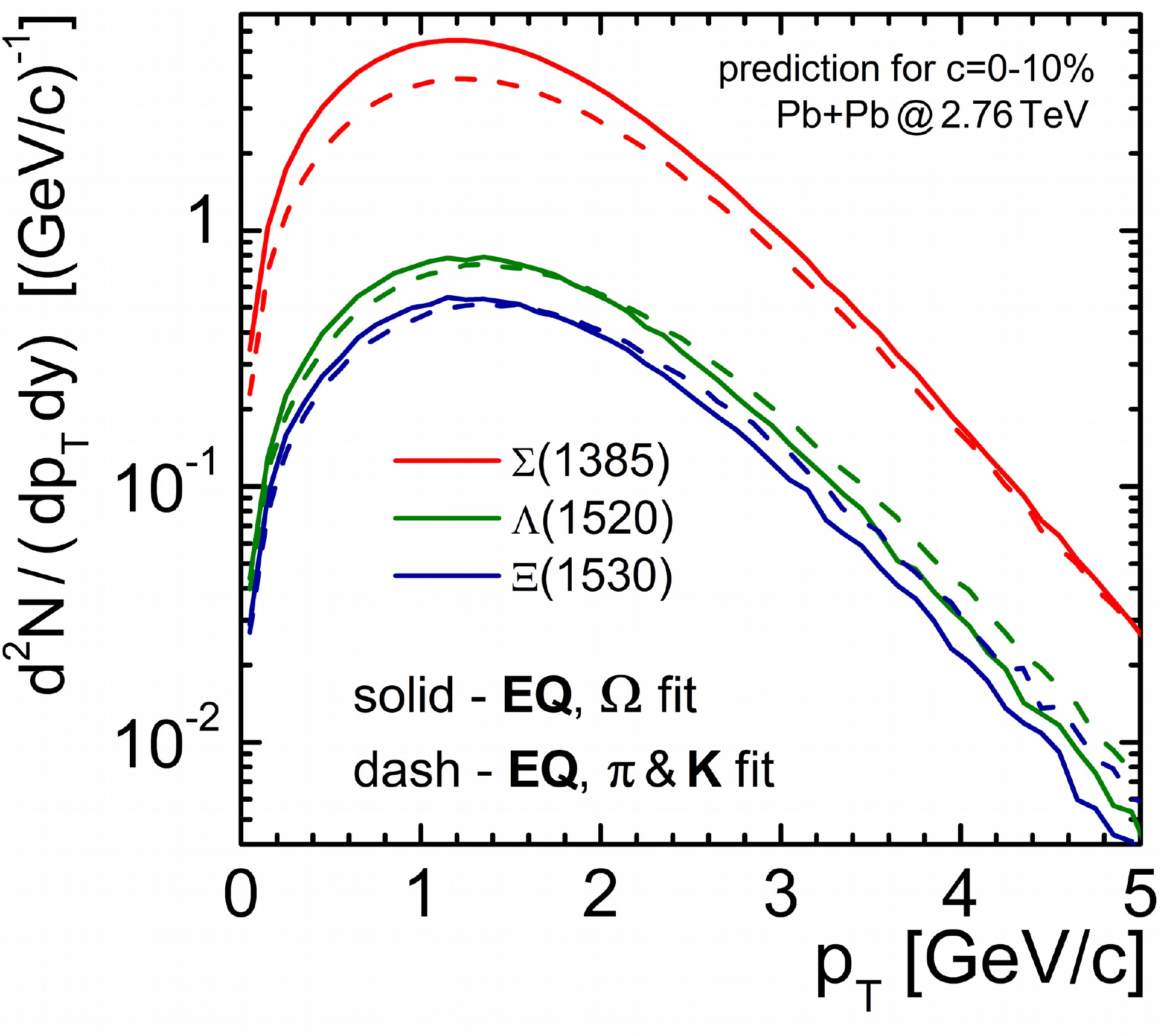}\vspace{-0.1cm}
\caption{Prediction for the half-sum of $\Sigma^+(1385)$, $\Sigma^0(1385)$, $\Sigma^-(1385)$, and their anti-particles, as well as $(\Lambda^0(1520)+\overline{\Lambda^0}(1520))/2$, and $(\Xi^0(1530)+\overline{\Xi^0}(1530))/2$.}
\vspace{-0.2cm}\label{fig-6}  
\end{figure*}

A good test for the non-equilibrium single freeze-out scenario \cite{Begun:2013nga,Begun:2014rsa} is the comparison to different resonances, especially strange resonances, because this scenario requires a special relation between the strange and the non-strange chemical potentials, depending on the quark content of a resonance. The heavy $\Lambda$, $\Xi$ and $\Omega$ can be still described by the non-equilibrium very well, if one assumes a smaller slope for them \cite{Begun:2014rsa}. This introduces the dependence on the mass of the resonance, but is also supported by smaller flow of heavy particles in other approaches, see e.g.~\cite{Melo:2015wpa}.
%
The parameters obtained in the fit to the 2.76 TeV Pb+Pb LHC data in equilibrium (EQ), non-equilibrium (NEQ)~\cite{Begun:2013nga,Begun:2014rsa}, and non-equilibrium with the possibility of pion Bose-Einstein condensation (BEC) on the ground state~\cite{Begun:2015ifa,Begun:2016cva} in hadron-resonance gas, using correspondingly modified SHARE~\cite{Petran:2013dva} and THERMINATOR~\cite{Chojnacki:2011hb} codes, are shown in Fig.~\ref{fig-1}.
One can see that the system is closer to the scenario with the condensate in central collisions. However, the uncertainty is rather large, which means that more mean multiplicities are needed to constrain the fit. At large chemical potentials finite size effects should be taken into account. 
The corresponding BEC fit of pion and kaon spectrum gives a good description of protons, while protons in EQ require a different freeze-out hypersurface. The amount and spectra of $\rho^0$ and $\eta$ mesons are significantly different in EQ and BEC, see Fig.~\ref{fig-2}.
Charged pions favor BEC\footnote{According to my best knowledge, there is no other model that explained this low $p_T$ excess of pions together with proton spectra without giving pions and protons extra parameters since the data appearance in 2012.}, see Fig.~\ref{fig-3}, while the $\eta/\pi^0$ ratio favors EQ (data from A.~Morreale~\cite{Morreale:2016dli}). However, the uncertainty, again, seems to be too large to judge.
Both BEC and EQ explain $K^0_S$ and $\phi$ spectra similarly good, see Fig.~\ref{fig-4}. The $K^*(892)^0$ is closer to BEC prediction. Note, that $K^*(892)^0$ was not included neither in the fit of mean multiplicities, nor in the fit of spectra (data from~\cite{Adam:2017zbf}). It means that BEC can be treated as an effective parameterizations of the freeze-out.
Strange baryons require different freeze-out hypersurface compared to that one for $\pi$, $K$, $p$, $K^0_S$, $K^*$, and $\phi$, see Fig.~\ref{fig-5}, and also~\cite{Melo:2015wpa,Chatterjee:2014lfa}. There is the mass dependence in BEC - the heavier the baryon, the smaller is the slope, i.e. the flow, or, equivalently, smaller radius of the hypersurface.
BEC predicts similar multiplicities and spectra of $\Lambda$(1520) and $\Xi$(1530), see Fig.~\ref{fig-6}. EQ predicts larger multiplicity difference between $\Lambda$(1520) and  $\Xi$(1530) than BEC. There is a significant dependence of the spectra on the freeze-out hypersurface for heavy strange baryons in BEC, while in EQ only $\Sigma$(1385) is sensitive to the hypersurface, see Figs.~\ref{fig-5},~\ref{fig-6}.

Therefore, one may conclude that $\pi$, $K$, $K^0_S$ and $\phi$ particles may have a common freeze-out hypersurface in both BEC and EQ models. The BEC additionally allows to explain protons, low $p_T$ pions, and $K^*(892)^0$. Strange baryons require different freeze-out in both models. The predictions for $\rho^0$, $\eta$, $\Sigma$(1385), $\Lambda$(1520), and $\Xi$(1530) are significantly different in BEC and EQ.
\\
\\
{\bf Acknowledgments:} I thank L.~Leardini, A.~Marin, C.~Markert and A.~Morreale for discussions. The research was partly carried out in laboratories created under the project "Development of research base of specialized laboratories of
public universities in Swietokrzyskie region" no POIG 02.2.00-26-023/08 dated May 19, 2009.

%
%
%

\end{document}